\documentclass[12pt]{amsart}
\usepackage{amsmath}

\newtheorem{thm}{Theorem}[section]
\newtheorem{lem}[thm]{Lemma}
\newtheorem{prop}[thm]{Proposition}

\numberwithin{equation}{section}

\newcommand{\Z}{{\mathbb Z}} 
\newcommand{\Q}{{\mathbb Q}}
\newcommand{\R}{{\mathbb R}}

\newcommand{\supp}{\operatorname{supp}}
\newcommand{\intinf}{\int_{-\infty}^\infty}

\newcommand{\meas}{\operatorname{meas}}

\newcommand{\ave}[1]{\left\langle#1\right\rangle} 

\newcommand{\Wf}{N_{f,\LH}} 

\renewcommand{\mod}{\;\operatorname{mod}}

\newcommand{\Prob}{{\mathbb{P}}}
\newcommand{\I}{1\!\!1}
\renewcommand{\Re}{{\mathfrak{Re}}}

\renewcommand{\i}{{\mathrm{i}}}
\renewcommand{\d}{{\mathrm{d}}}
\renewcommand{\^}{\widehat}

\newcommand{\aveT}[1]{\left\langle#1\right\rangle_{\weight,T}} 
\newcommand{\Wos}{S_{f,\LH}}

\newcommand{\weight}{w}

\newcommand{\HH}{\mathbb H}

\newcommand{\area}{\operatorname{area}} 
\newcommand{\tr}{\operatorname{tr}}

\newcommand{\amp}{\beta} 
\newcommand{\norm}{\mathcal N} 

\newcommand{\meanf}{\overline{\Wf}} 
\newcommand{\main}{M}

\newcommand{\Leg} [2]{\left( \frac{#1}{#2} \right)} 

\newcommand{\width}{T} 

\newcommand{\sign}{\eta} 

\newcommand{\crad}{\pi} 

\newcommand{\rem}{\mathcal E} 

\newcommand{\LH}{L} 
\newcommand{\LM}{\mathcal L} 

\newcommand{\llp}{\prec\prec} 
\newcommand{\ggp}{\succ\succ} 

\begin{document}

\title[CLT for the spectrum of the modular domain] 
{A  Central Limit Theorem for the spectrum of the modular  domain} 
\author{Ze\'ev Rudnick}
\address{Raymond and Beverly Sackler School of Mathematical Sciences,
Tel Aviv University, Tel Aviv 69978, Israel
({\tt rudnick@post.tau.ac.il}) 
}

\date{November 10, 2004} 

\thanks{A version of this paper was presented in the IAS/Park City
Mathematics Institute summer session on Automorphic Forms and
Applications  in July 2002 as part of the author's mini-course on
Arithmetic Quantum Chaos.  
Supported by a grant from the Israel Science Foundation, founded by
the Israel Academy of Sciences and Humanities and  a Leverhulme Trust
Linked Fellowship at Bristol University.}

\maketitle


The statistics of the high-lying eigenvalues of the Laplacian on
a Riemannian manifold have been intensively studied in the past few
years by physicists working on ``quantum chaos''.   
A number of fundamental insights have emerged from these studies,
though to date these have yet to be set on rigorous footing. In the
case the manifold at hand is of arithmetic origin, these studies are
related to some profound number theoretical problems and as such may
be more amenable to investigation. 
In this note I make use of the arithmetic structure of the modular
domain to establish Gaussian fluctuations in its spectrum for certain
smooth counting functions. 

\tableofcontents 

\section{Background} 
To set the stage, I start with describing some of what is currently believed
to hold for the statistics of the eigenvalues. 
Given a compact Riemannian surface $M$, 
Weyl's law for the eigenvalues $E_j$  of the
Laplacian says that the  number of eigenvalues below $x$  grows 
linearly with $x$:  
$$
\#\{E_j \leq x \} \sim \frac{\area(M)}{4\pi}  x, \quad  \mbox{ as }  x\to \infty \;.
$$ 

Let $n(E,\LM)$ be the number of levels in a window around $E$ for
which the leading term in Weyl's law predicts $\LM$ levels:  
$$
n(E,\LM) = \#\{E- \frac {2\pi }{\area(M)}\cdot \LM <E_j 
<E+\frac {2\pi}{\area(M)}\cdot \LM \} 
$$
and more generally for a test function $f$ define 
\begin{equation}\label{eigenvalue n_f} 
n_f(E,\LM) := \sum_j f(\frac{\area(M)}{4\pi} \cdot \frac{(E_j-E)}{\LM}) 
\end{equation}
which counts the levels lying in a ``soft'' window of length
$\frac{4\pi}{\area(M)} \LM$ about
$E$. In the above  $\LM=\LM(E)$ depends on the 
location $E$. In what follows we will usually write $n(\LM)$ for
$n(E,\LM)$, the dependence on $E$ implicitly understood. 

To study the statistical behaviour of $n(\LM)$  we need to consider $E$
as random, drawn from a certain distribution on the line. 
We denote by $\ave{\cdot }$ this kind of energy averaging, 
e.g. $\ave{F} =\frac 1E \int_{E}^{2E}F(E')dE'$.   
Weyl's law leads us to expect that the mean value 
of $n(\LM)$ is $\LM$ and likewise that of $n_f(\LM)$ is 
$\LM\cdot \int f(x)dx$.  

\subsection{Number variance} 
The variance of $n_f(E,\LM)$ from its expected value  is: 
$$
\Sigma_f^2(E,\LM) = \ave{|n_f(\LM) - \ave{n_f(\LM)}|^2} 
$$
It is customary to express the number variance by means of an integral
kernel $K_E(\tau)$, called the ``form factor'', 
so that as $E\to \infty$  
$$
\Sigma^2_f(E,\LM) \sim \LM\cdot \intinf \^f(u)^2 K_E(\frac {u}{\LM}) du  =
\intinf (\LM\^f(\LM\tau))^2 K_E(\tau) d\tau \;.
$$
where $\^f(y) = \intinf f(x)e^{-2\pi i xy} dx$ is the Fourier
transform of $f$.

For ``generic'' surfaces, Berry  \cite{Berry85, Berry86} argued 
that as $E\to \infty$, 
the behaviour of $\Sigma^2_f(E,\LM)$ for $\LM$ in the
range\footnote{The symbol $f(x)\llp g(x)$ means that 
$f(x)/g(x) \to 0$} 
\begin{equation}\label{universal range} 
1\llp \LM \llp \LM_{max}= \sqrt{E}   
\end{equation}
is {\em universal}, 
depending only on the coarse dynamical nature  of the geodesic 
flow on the surface, and follows that of one of a small number of
random matrix ensembles: 
If the flow is
{\em integrable} (as in the case of a flat torus)  
then $\Sigma^2_f(E,\LM) \sim \LM\cdot \intinf f(x)^2 dx$ for $\LM\to \infty$, 
as in the Poisson model of uncorrelated levels.  
If the flow is {\em chaotic} (as in the case of negative curvature) 
then the behaviour is as in the Gaussian Orthogonal Ensemble (GOE):  
For the sharp window ($f=\mathbf 1_{[-1/2,1/2]}$), this is given by  
$\Sigma^2(E,\LM) \sim  \frac 2{\pi^2}\log \LM$ for $\LM\to \infty$.   
For sufficiently smooth $f$, in the GOE we have 
$\Sigma^2_f(E,\LM)\sim 2\intinf \^f(u)^2 |u| du$, 
that is the variance of sufficiently smooth statistics tends to a finite
value as $\LM\to \infty$.  
The form factors  for the random models are  $K^{pois}(\tau)
\equiv 1$,  and 
$$
K^{GOE}(\tau)
=\begin{cases} 2|\tau|-|\tau|\log(1+2|\tau|),& |\tau|\leq 1 \\  
2-|\tau| \log \frac{1+2|\tau|}{2|\tau|-1}, & |\tau|>1 \end{cases} \;.
$$

It is to be emphasized that the above behaviour is only valid in the
universal regime $1\llp \LM\llp \sqrt{E}$; for $\LM\ggp \sqrt{E}$ the
integrable case  is fairly well understood (at a rigorous level), 
see the survey \cite{Blehersurvey}: The
variance grows as $\sqrt{E}$ (a classical result \cite{Cramer} in the
case of the standard flat torus). In the chaotic case it is believed
\cite{Berry85, Berry86} that generically, the
number variance continues to be small as in the universal regime.

\subsection{Fluctuations} 
Our main interest here is in the value distribution  of the
normalized linear statistic 
$$ \frac{n_f(E,\LM) -\ave{n_f(\LM)}}{\sqrt{\Sigma^2_f(E,\LM)}}
$$ 
as $E$ varies. 
In all the statistical models (Poisson and GOE/GUE), it is a standard 
Gaussian \cite{Politzer, DS, CL}. 

In the integrable case, when $\LM\ggp \sqrt{E}$, the distribution is
known (\cite{H-B}, \cite{Blehersurvey}), and is definitely not
Gaussian. Inside the universal regime \eqref{universal range}, 
the distribution is believed to be Gaussian in both
the integrable \cite{Ble-Leb} and chaotic \cite{ABS, bs, Steiner} cases. 
In the special case of the standard flat torus, this has been proved
in a small part of the universal regime near $\sqrt{E}$ \cite{HR}.

\subsection{The modular domain} 
We start with the upper half-plane  $\HH=\{x+\sqrt{-1} y:y>0\}$
equipped with  the hyperbolic metric $ds^2 = y^{-2}(dx^2 +dy^2)$,
which has constant curvature equal $-1$.  
The  Laplace-Beltrami operator for this metric is given by 
$\Delta = 
y^2(\frac{\partial^2}{\partial x^2}+\frac{\partial^2}{\partial y^2})$. 
The orientation-preserving isometries of the metric $ds^2$ are the
linear fractional transformations $PSL(2,\R) =SL(2,\R)/\{\pm I\}$.

The modular domain is the Riemann surface obtained by identifying
points in the upper half plane which
differ by a linear fractional transformation with integer
coefficients, that is by elements of the modular group
$$
\Gamma:=PSL(2,\Z) = SL(2,\Z)/\{\pm I\} \;. 
$$ 
The resulting surface $\HH/\Gamma$ is non-compact and has cone points,
but has finite hyperbolic area: $\area(\HH/\Gamma) = \pi/3$.

The spectrum of the Laplacian on the modular domain 
has a continuous component. 
Nonetheless, Selberg \cite{Selberg} showed that 
a version of Weyl's law holds for the discrete spectrum: 
If we write the eigenvalues of $-\Delta$ on $L^2(\HH/\Gamma)$ 
in the form $E_j=1/4+r_j^2$, then: 
\begin{equation}\label{selberg-weyl}
\#\{r_j \leq T\} =\frac{ \area(\HH/\Gamma) }{{4\pi}}T^2 
 -\frac 2\pi T\log T +c_1  T +
O(\frac T{\log T}) 
\end{equation} 
where $c_1=\frac {2+\log \frac{\pi}2}\pi$.

In the case of the modular domain, deviations from generic statistics
were discovered \cite{BGS, BGGS, BSS}. 
Although the geodesic flow is chaotic, the
local statistics of the spectrum seem Poissonian, 
and Bogomolny, Leyvraz and Schmit \cite{bls} argued that
the behaviour of the form factor is given by 
$$
K_E(\tau) \sim \begin{cases} 
c_1 \frac{\exp(c_2 \sqrt{E}\tau)}{\sqrt{E}}, & 
\frac 1{\sqrt{E}} \llp \tau \llp \frac{\log\sqrt{E}}{\sqrt{E}} \\ 
1, & \tau \ggp  \frac{\log\sqrt{E}}{\sqrt{E}}
\end{cases}
$$
for some constants $c_1,c_2>0$, 
that is to say, in the universal regime we have 
$$
\Sigma^2_f(E,\LM) \sim \begin{cases} 2c_1 \frac{\LM}{\sqrt{E}}
\int_0^\infty \^f(u)^2 \exp(c_2 \frac{\sqrt{E}u}{\LM}) du,  &
\frac{\sqrt{E}}{\log E} \llp \LM \llp  \sqrt{E} 
\\ 
\LM \cdot \intinf \^f(u)^2 du , & 1\llp \LM \llp \frac{\sqrt{E}}{\log E}
\end{cases}
$$

The only rigorous results known  concern the  the
closely related case where the modular group is replaced by quaternion
groups: In the 1970's Selberg \cite[Chapter 2.18]{Hejhal} gave a {\em lower}
bound for the variance $\Sigma^2(E,E)$ of  the sharp counting function
$n(E)$ of the form 
$\Sigma^2(E,E) \gg \sqrt{E}/(\log E)^2$. 
Luo and Sarnak \cite{LS} gave lower bounds for the
averaged number variance of the sharp counting function $n(E,\LM)$:  
$$\frac 1{\LM}\int_{0}^{\LM}\Sigma^2 (E,\LM')d\LM'\gg
\sqrt{E}/(\log E)^2, \qquad   \sqrt{E}/\log E
\llp \LM \llp \sqrt{E}
$$ 
in the case of  arithmetic (co-compact) groups. 
In the case of the hard window $f=\mathbf 1_{[-1/2,1/2]}$ no {\em
upper} bounds are currently available.


\section{Formulation of results} 
\subsection{Definition of the smooth counting functions} 
Let $f$ be an even test function, whose Fourier transform 
$\^f\in C_c^\infty(\R)$ is smooth and compactly supported, and
normalized by requiring that 
$$ 
\intinf f(x) dx = 1,
$$
and  that 
$$
\sup\{|\xi|: \^f(\xi)\neq 0 \} = 1 \;. 
$$ 
 
The eigenvalues of the Laplacian are parametrized by $E_j =
1/4+r_j^2$. 
We define smooth counting functions by 
\begin{equation}
\Wf(\tau) = \sum_{j\geq 0} f(\LH(r_j-\tau)) + f(\LH(-r_j-\tau))
\end{equation}
This in essence carries the same information as 
\eqref{eigenvalue n_f}.  
The relation between the expected number of levels $\LM $ and the 
inverse width $\LH$ of the momentum window is 
$$
\LM = \frac{\area(\HH/\Gamma) \sqrt{E}}{2\pi \LH}  =
\frac{\sqrt{E}}{6\LH} \;.
$$

The leading order behaviour of $\Wf$ is given by 
$$
\meanf(\tau) : = \frac 1{2\pi} \intinf \{ f(\LH(  r-\tau))+ f(\LH( -
r-\tau)) \}  \main(r) \d r  
$$
where  
$$
\main(r):= \frac{\area(  \HH / \Gamma )}{2} r\tanh(\pi r) - 
\frac{\Gamma'}{\Gamma}(1+\i r)  - 
\frac{\Gamma'}{\Gamma}(\frac 12+\i r) \;. 
$$
(In keeping with tradition, I use the symbol $\Gamma$ 
for both the modular group and the Gamma function).  
The term $\meanf(\tau)$ is  asymptotic to $\LM$: 
$$
\meanf(\tau) \sim 
2 \frac{ \area(\HH/\Gamma)}{{4\pi}}\intinf f(x)\d x \frac{\tau}\LH  +
O(\frac {\log \tau}{\LH} ) \sim 
\frac 16 
\frac{\tau}\LH \;. 
$$

\subsection{The results} 
We will see that  $\Wf-\meanf$ has mean zero 
and show that  the variance of $\Wf$,  
when $\limsup \pi \LH / \log T <1$, is asymptotic to 
\begin{equation}\label{varfor SL2} 
\sigma_{\LH}^2:
=\frac{2\kappa}{\pi \LH}\int_0^\infty \^f(u)^2 e^{\pi \LH u}\d u
\end{equation}
where 
$
\kappa= \frac{1015}{864}\prod_{p\neq 2} 
(1+\frac{p^4-2p^3+1}{(p^2-1)^3}) 
= 1.328\dots  
$.  
Thus when the expected number of levels $\LM$ satisfies 
\begin{equation*}\label{our variance regime} 
\area(\HH/\Gamma)  \frac{\sqrt{E}}{\log E} < \LM \llp T=\sqrt{E} \;.
\end{equation*}
the form factor $K_E(\tau)$ is given by 
$$
K_E(\tau) =  c_1\frac {\exp c_2 \sqrt{E} \tau}{\sqrt{E}} 
$$
with $c_1 = 6\kappa/\pi$, $c_2 =\pi/6$.


Our main result is that  the fluctuations of $\Wf$ are Gaussian: 
\begin{thm}\label{main thm}  
Assume that  $\LH\to \infty$ as $T\to \infty$ but $\LH=o(\log T)$. Then 
the limiting value distribution of $(\Wf-\meanf)/\sigma_{\LH}$ 
is  a standard Gaussian, that is 
$$
\lim_{T\to \infty} 
\frac 1T \meas \{ \tau \in [T,2T]: 
\frac{\Wf(\tau)-\meanf(\tau)}{\sigma_{\LH}} <x  \} = 
\int_{-\infty}^x e^{-u^2/2} \frac{\d u}{\sqrt{2\pi}} 
$$
\end{thm}

The reason that we need to assume that $\LH = o(\log T)$ is that we
prove Theorem~\ref{main thm} by the method of moments, and in
computing the $K$-th moment we find Gaussian moments for $\LH < c_K
\log T$, where $c_K\to 0$ as $K$ grows.

\noindent{\bf Plan of the paper:} 
In sections~\ref{sec:modular group} and \ref{sec:length spectrum} we
give some results on the hyperbolic conjugacy classes of the modular
group. In section~\ref{sec: An expansion}  
we use Selberg's trace formula to express $\Wf(\tau)-\meanf(\tau)$ as
a sum $\Wos(\tau)$ 
over hyperbolic conjugacy classes plus a negligible term. 
The variance of $\Wos$ is computed in section~\ref{mean and variance}  
and the higher moments in section~\ref{sec:\Wos is gaussian}.   
We prove Theorem~\ref{main thm} in section~\ref{sec:conc}.

\section{The modular group}\label{sec:modular group}
\subsection{Conjugacy classes} 
To analyze  $\Wf$ we use the Selberg trace formula, which for a 
discrete co-finite subgroup $\Gamma\subset PSL(2,\R)$ 
relates a sum over the spectrum of the Laplacian on $L^2(\HH/\Gamma)$
with a sum over the conjugacy classes of the group $\Gamma$. 
We review some background material on these classes for the modular
group $PSL(2,\Z)$. 


The conjugacy classes  are divided into the class which
consists of the identity element, hyperbolic, elliptic and parabolic
classes. 
The  hyperbolic conjugacy classes in $\Gamma$ are represented by
matrices $P$ which are diagonalizable over the reals and are conjugate
to a matrix of the form $\begin{pmatrix} \lambda & \\ &\lambda^{-1}
\end{pmatrix}$ with $\lambda>1$. 
The {\em norm}  of $P$ is defined as $\norm(P)=\lambda^2$. The norm is 
therefore related to the trace of the corresponding group element by  
$$
\norm(P)^{1/2} + \norm(P)^{-1/2} = |\tr(P)|
$$
We can write as each such $P$ as $P=P_0^k$ where $P_0$ is 
{\em primitive} and $k\geq 1$, where an element of $\Gamma$  is
primitive if it cannot be written as an essential power of another
element.  As is well known, primitive hyperbolic conjugacy classes
correspond to closed geodesics on the Riemann surface $\HH/\Gamma$.

In the case of the modular group $\Gamma=PSL(2,\Z)$, the traces are
integers. If $P$ is  
a hyperbolic class with trace  $|\tr(P)| = n$, $n>2$ then its norm is   
\begin{equation}\label{norm}
\norm(n)  = ( \frac{n+\sqrt{n^2-4}}2 )^2
\end{equation}

For the modular group,   {\em primitive}
hyperbolic conjugacy classes are parametrized by indefinite binary
quadratic forms as follows (cf.  \cite{Sarnak82}): 
Take a binary quadratic form  $Q_{a,b,c}(x,y):=ax^2+b xy+cy^2$, with
$a,b,c\in\Z$.  
The discriminant of $Q_{a,b,c}$ is $d:=b^2-4ac$. The form
$Q_{a,b,c}$  is indefinite iff $d>0$. We assume that $d$ is not a
perfect square. 
We say that $Q_{a,b,c}$ is primitive if
$\gcd(a,b,c)=1$. 
Two binary quadratic forms $Q$, $Q'$ are equivalent if $Q'(x,y) = Q(ax+by,
cx+dy)$ for an element $\gamma = \begin{pmatrix} a&b\\ c&d \end{pmatrix}$ of
$SL(2,\Z)$;   since the forms are quadratic, they are also equivalent
under $-\gamma$ and hence equivalence is over $PSL(2,\Z)$. 
Let $h(d)$ be the number of equivalence classes of
primitive binary quadratic forms of discriminant $d$. 

The automorphs of $Q_{a,b,c}$ are all of the form 
$$
\pm P(t,u)=
\pm \begin{pmatrix} \frac 12(t-bu)& -cu\\ au& \frac 12(t+bu) \end{pmatrix}
$$
where $(t,u)$ solve the Pellian equation 
\begin{equation}\label{Pell}
t^2-du^2=4
\end{equation}
If $u\neq 0$ then these are hyperbolic elements of $SL(2,\Z)$ with
norm $\norm(P) = (t+u\sqrt{d})^2$ and trace $t$. 
 
Let $\epsilon_d = \frac 12(t_d+u_d\sqrt{d})$ ($t_d$, $u_d>0$) 
be the fundamental solution of \eqref{Pell}. 
Then the matrix $P(t_d,u_d)$ is a primitive
hyperbolic matrix $P_0$ of trace $\tr(P_0)=t_d$ and norm $\norm(P_0) =
\epsilon_d^2$. It turns out that in this way we get a bijection
between equivalence classes of primitive binary quadratic forms and
conjugacy classes of primitive hyperbolic matrices in $PSL(2,\Z)$. 
Thus the number of primitive hyperbolic conjugacy
classes of norm $ \epsilon_d^2$ is precisely the class number $h(d)$.

\subsection{The amplitude $\amp(n)$} 

We define, for $n>2$,  
\begin{equation}\label{the amplitude}
\amp(n) := \frac 12  
\sum_{ \tr(P)=n} 
\frac{\log \norm(P_0)}{\norm(P)^{1/2}-\norm(P)^{-1/2}}
\end{equation}
the sum over all conjugacy classes $\{P\}$ in $PSL(2,\Z)$  with $|\tr(P)|=n$, 
equivalently with norm $\norm(n)$ given by \eqref{norm}. 
%
These quantities turn out to be crucial in our analysis. 
The factor $1/2$ in the definition is inserted among other reasons to
give numbers with mean value $1$ (as can be seen from the Prime
Geodesic Theorem): 
$$
\sum_{n\leq N} \amp(n) \sim N, \mbox{ as } N \to \infty \;.
$$

As representatives of the conjugacy classes of matrices with trace
$n>2$ we can take the matrices $P_0^k =P(t_d,u_d)^k=P(n,u)$ where $d$
runs over all discriminants, $k\geq 1$ and $n^2-du^2=4$, $n>2$, 
$u\geq 1$.  
Thus we see that 
\begin{equation}\label{amp(n) in terms of class numbers}  
\amp(n) = \sum_{\substack{d,u\geq 1\\  n^2-du^2=4}}
\frac{h(d)\log \epsilon_d}{\sqrt{d u^2} } \;.
\end{equation}

Dirichlet's class number formula allows us to use 
\eqref{amp(n) in terms of class numbers}  to express $\amp(n)$ in
terms of Dirichlet $L$-functions:
For a discriminant $d$ on associates the quadratic character  $\chi_d$ 
given by $\chi_d(p)=\Leg{d}{p}$ for $p$ an odd prime, $\chi_d(2)=1$ if
$d\equiv 1\mod 8$, $\chi_d(2)=-1$ if $d\equiv 5 \mod 8$ and
$\chi_d(-1)=1$.  The associated $L$-function is $L(s,\chi_d) =
\sum_{n\geq 1} \chi_d(n) n^{-s}$, $\Re(s)>1$.  
Dirichlet's class number formula is 
$$
h(d)\log \epsilon_d =\sqrt{d}L(1,\chi_d) \;. 
$$
Inserting the class number formula into 
\eqref{amp(n) in terms of class numbers}   
we find that 
\begin{equation}\label{amp(n) in terms of L-functions} 
\amp(n) = \sum_{\substack{d,u\geq 1: du^2=n^2-4}} \frac 1u L(1,\chi_d) \;.
\end{equation} 
As a consequence, one can get an upper bound of 
$\amp(n) =O( (\log n)^2)$
 by using $L(1,\chi_d)\ll \log d$. 
What is more useful to us is that, in the mean square, $\amp(n)$ is
constant: 
\begin{lem}[M. Peter \cite{Peter}]\label{Peter's formula}
\begin{equation}
\sum_{n\leq N} \amp(n)^2 \sim \kappa N,\qquad N\to\infty
\end{equation}
where $\kappa$ is given by the product over primes 
\begin{equation}
\label{Peter's constant}
\kappa= \frac{1015}{864}\prod_{p\neq 2} 
(1+\frac{p^4-2p^3+1}{(p^2-1)^3}) 
= 1.328\dots  
\end{equation}
\end{lem}
This (complicated) expression was derived heuristically by
Bogomolny, Leyvraz and Schmit \cite{bls} and  
proven by Manfred Peter \cite{Peter}, who uses the expression 
\eqref{amp(n) in terms of L-functions} 
for $\amp(n)$ in terms of $L(1,\chi)$ and methods related to work
on moments of class numbers \cite{Barban}.  
For an extension to the case of congruence groups, see \cite{Lukianov}.

\section{The length spectrum}\label{sec:length spectrum}


We will need to study alternating sums 
of the form 
$$
\sum_{j=1}^K \pm \log \norm(n_j)
$$
The first question is to when these alternating sums vanish. 

We say that a relation 
\begin{equation}\label{relation}
\sum_{j=1}^K \sign_j \log \norm(n_j)=0,\qquad \sign_j=\pm 1
\end{equation}
is {\em non-degenerate} if no sub-sum vanishes, that is if there is no
proper subset $S\subset\{1,\dots,K\}$ for which $\sum_{j\in S} \sign_j
\log\norm(n_j) = 0$. The existence of non-degenerate
relations \eqref{relation} forces severe constraints. To explain
these, recall that $\norm(n)$ is 
a unit in the real quadratic field $\Q(\sqrt{n^2-4})$. We claim that 
such  such relations can occur only if all these units lie in the same
quadratic field.

\begin{lem}\label{lem:Indep}
Let $\sum_{j=}^K \pm  \log \norm(n_j)=0$ be a non-degenerate
relation. Then all the norms $\norm(n_i)$ lie in the same quadratic
field, that is for some common $d$ we have  $n_i^2-4=df_i^2$ for all
$i$. 
\end{lem}

%
%

\begin{proof}
We can write each norm as a power of the fundamental unit of the
quadratic field in which it lies. Thus it will suffice to show 
that if  $F_1,\dots, F_K$ be distinct real quadratic fields, then  
the fundamental units $\epsilon_i$ of $F_i$ 
are multiplicatively independent. 

Let $E=F_1\vee \dots \vee F_K$ be the compositum of the fields
$F_i$. This is a Galois extension of the rationals with Galois group
$G=Gal(E/\Q)$ an  elementary Abelian $2$-group $(\Z/2\Z )^s$, for some
$s\leq K$.  If we denote by $U_E$ the unit group of $E$, then $G$
acts on $U_E$ and hence we get a linear representation on the vector
space $\Q\otimes U_E$.

We claim that the $\epsilon_i$ are eigenvectors of $G$,
that is $\sigma(\epsilon_i) = \epsilon_i^{\chi_i(\sigma)}$ for all
$\sigma \in G$, where 
$\chi_i:G\to \{\pm 1\}$ are {\em distinct} characters. 
This forces them to be multiplicatively independent.

Indeed, since we have an Abelian extension, all subfields are Galois
and in particular $F_i$ are preserved by $G$. Since the unit group is
also  preserved this means that under the action of any element
$\sigma\in G$, $\epsilon_i$ is taken to a unit of $F_i$
which is necessarily $\epsilon_i^{\pm 1}$. That is we have a character
$\chi_i$ of $G$ with $\sigma(\epsilon_i) =
\epsilon_i^{\chi_i(\sigma)}$. The characters $\chi_i$ are distinct
since the kernel of $\chi_i$ is precisely $Gal(E/F_i)$. 
\end{proof}


We next get a lower bound for $\sum_{j=1}^K \pm
\log \norm(n_j)$ in the case it is non-zero.

\begin{lem}\label{use Liouville}
i) If $m\neq n$ then 
$$
|\log \norm(m)-\log \norm(n)| \gg  
 \frac 1{\min(m,n)} \;.
$$

ii) Suppose $ \sum_{j=1}^K \pm \log \norm(n_j)$ is nonzero. Then 
$$
|\sum_{j=1}^K \pm \log \norm(n_j) | \gg  
\frac 1 {\left(\prod_{j=1}^K \norm(n_j) \right)^{2^{K-1}-1/2}}
$$
\end{lem}
\begin{proof}
i) Indeed, since $\log \norm(n)=2\log n+O(1/n^2)$,
if $m\neq n$, say $m>n$, then 
$$
\log \norm(m)-\log \norm(n)= 
2\log \frac{m}{n} +O(\frac 1{n^2}) \;. 
$$ 
Since $\log \frac mn \geq \log \frac{n+1}n \gg \frac 1n$, we find 
$$
\log \norm(m)-\log \norm(n)\gg \frac 1n  = \frac 1{\min(m,n)} \;.
$$

ii) Let $\lambda_j = \frac {n_j+\sqrt{n_j^2-4}}2$ so that $\norm(n_j) = \lambda_j^2$, and set 
$\alpha = \prod_{j=1}^K \lambda_j^{\pm 1}$. 
If $|\alpha-1|\leq 1/2$ then  
$$ 
|\sum_{j=1}^K \pm \log \norm(n_j) |=2|\log \alpha| \gg |\alpha-1|
$$ 
%

So it suffice to give a lower bound for $|\alpha-1|$, assuming
$\alpha\neq 1$. This follows from Liouville's theorem on Diophantine
approximation of algebraic numbers by rationals; we give an explicit
proof as follows: Let $E = \Q(\sqrt{n_1},\dots, \sqrt{n_K})$, which
is a Galois extension of the rationals with Galois group
$G=Gal(E/\Q)$ which is an elementary abelian $2$-group of order
$2^s$, for some $s\leq K$. Moreover $\alpha\in E$ is an algebraic
integer, and hence the norm $\norm_{E/\Q}(\alpha-1)$ is a nonzero
rational integer, hence has absolute value at least $1$. 
Thus 
$$
|\norm_{E/\Q}(\alpha-1)| = |\alpha-1| \prod_{id\neq \sigma\in G} |\alpha^\sigma -1| \geq 1
$$
%

Since $\lambda_j^\sigma = \lambda_j^{\pm 1}$ for all $\sigma \in G$, we have 
$$
|\alpha^\sigma-1| \leq \prod_{j=1}^K \lambda_j +1
$$
Thus
$$
|\alpha-1| \geq \frac 1{ (\prod_{j=1}^K \lambda_j +1)^{|G|-1}}\gg 
\frac 1{(\prod_{j=1}^K \norm(n_j))^{(2^K-1)/2}}
$$
\end{proof} 


\section{ An expansion for $\Wf$}\label{sec: An expansion}

\subsection{The Selberg Trace Formula} 
We will transform $\Wf$ by using the Selberg trace
formula \cite{Selberg}: 
Let $g\in C_c^\infty(\R)$ be a an even, smooth and compactly supported  function, 
and let 
$$
h(r)=\intinf g(u)e^{\i ru}\d u
$$ 
so that 
$$
g(u)= \frac 1{2\pi}\intinf h(r)e^{-\i ru}\d r \;.
$$ 
The Selberg Trace Formula for a discrete co-compact sub-group
$\Gamma\subset PSL(2,\R)$  with no elliptic
elements is the identity \cite{Selberg} 
\begin{equation}\label{selberg trace formula}
\begin{split}
\sum_{j\geq 0} h(r_j)  
&= 
\frac{\area(\HH/\Gamma)}{4\pi} \intinf h(r) r\tanh(\pi r) \d r \\ 
&+ \sum_{\{P\} \text{ hyperbolic}} 
\frac{\log \norm(P_0)}{\norm(P)^{1/2}-\norm(P)^{-1/2}} g(\log \norm(P)) \\
\end{split}
\end{equation} 
where the sum is over all hyperbolic conjugacy classes of $\Gamma$. 

In the case of the modular group, the hyperbolic terms can be written as 
\begin{equation}\label{hyperbolic terms}
2 \cdot \sum_{n>2} \amp(n) g(\log \norm(n)) 
\end{equation} 
where the amplitude $\amp(n)$ is given by \eqref{the amplitude}. 

For groups with elliptic elements, there is an extra contribution to
the RHS of \eqref{selberg trace formula}  
which is a  sum over the finitely many conjugacy
classes of elements $E$ of finite order $m\geq 2$:
\begin{equation}\label{elliptic terms} 
\sum_{\{E\}} \sum_{k=1}^m \frac{1}{m\sin(\pi k/m)} \intinf
h(r) \frac{e^{-2\pi kr/m}}{1+e^{-2\pi r}}\d r
\end{equation}

For discrete groups whose fundamental domain is non-compact but of
finite volume, that is with cusps, there are extra terms coming from
the contribution of the continuous spectrum and parabolic elements.
For $\Gamma=PSL(2,\Z)$, these terms are given explicitly by 
\cite{Hejhal2}: 
\begin{multline}\label{parabolic terms} 
g(0)\log \frac {\pi}{ 2} 
- \frac{1}{2\pi} \intinf h(r) \left( \frac{\Gamma'}{\Gamma}(1+\i r) +
\frac{\Gamma'}{\Gamma}(\frac 12+\i r) \right) \d r  \\
+ 2\sum_{n=1}^\infty  \frac{\Lambda(n)}{n} g(2\log n)
\end{multline}
where $\Lambda(n)$ is the von Mangold function. 


\subsection{Transforming $\Wf$} 
We now apply the trace formula to derive an alternative expression for
$\Wf$. 
Taking 
$$
h(r) = f(\LH(r-\tau)) +  f(\LH(-r-\tau))
$$ 
so that 
$$
g(u)  = \frac 1{2\pi \LH} \^f(\frac u{2\pi \LH})
\left( e^{-\i \tau u}+e^{\i \tau u} \right)
$$ 
we find that 
\begin{equation}
\Wf(\tau) =\overline{\Wf}(\tau) + \Wos(\tau) + \rem
\end{equation}
where: 
 
\noindent $\bullet$ The term $\meanf$ is given by the contribution of
the identity class to \eqref{selberg trace formula}  and part of the
parabolic terms in \eqref{parabolic terms}:  
$$
\meanf(\tau) = 
 \frac 1{2\pi} \intinf \{ f(\LH(  r-\tau))+f(\LH( - r-\tau)) \} \main(r)
\d r
$$
where 
$$\main(r) = 
\frac{\area( \HH/\Gamma )}{2} r\tanh(\pi r)  - 
\frac{\Gamma'}{\Gamma}(1+\i r)  - \frac{\Gamma'}{\Gamma}(\frac 12+\i r) \;.
$$
By Stirling's formula, we have
$$
\meanf(\tau) = \frac 16 \intinf f(x)\d x \frac{\tau}\LH + 
O(\frac{\log\tau}\LH) \;.
$$

\noindent $\bullet$ 
The term $\Wos(\tau)$ is the contribution
of the hyperbolic classes \eqref{hyperbolic terms}: 
\begin{equation}\label{sc \Wos} 
\Wos(\tau) =
\frac 1{\pi \LH}\sum_{n>2} \amp(n) 
\^f(\frac {\log \norm(n) }{2\pi \LH}) 
\left( e^{-\i \tau \log \norm(n)} +  e^{\i \tau \log \norm(n)}  \right)
\end{equation}
The sum \eqref{sc \Wos} contains only terms with 
$\log \norm(n)\leq 2\crad \LH$, that is $n<e^{\crad \LH}$.

As we will see below, 
it is the term $\Wos(\tau)$ which is responsible for the fluctuations of
$\Wf(\tau)$, and its variance is asymptotic to $\sigma_{\LH}^2$.  
As we can see from the formula \eqref{sc \Wos}, since $\^f$ has
compact support we have $\Wos(\tau) \equiv 0$ for $\LH\ll 1$. 

\noindent $\bullet$ 
$\rem$ is the contribution of the 
elliptic classes \eqref{elliptic terms}  
and the remaining part of the parabolic contribution \eqref{parabolic
terms}, namely 
\begin{equation}\label{residual part} 
\frac 1{\pi \LH} \^f(0) \log \frac{\pi}2 + \frac 1{\pi \LH}
\sum_{n=1}^\infty \frac{\Lambda(n)}{n}\^f(\frac{\log n}{\pi \LH})
2\cos(2\tau \log n). 
\end{equation}  
$\rem$  is easily seen to be negligible, that is 
$\rem =o(\sigma_{\LH})$. 
Indeed, the contribution of the elliptic elements 
is easily seen to be $O(e^{-\mbox{const.} \tau}/\LH)$. 
As for  \eqref{residual part}, 
this is bounded as $\LH\to \infty$ by (say) Mertens' theorem. Moreover
the mean value of \eqref{residual part}  clearly vanishes as $T\to
\infty$.

We thus see  that the difference between the centered counting function 
$\Wf(\tau)-\meanf $ and the  sum $\Wos(\tau)$ over hyperbolic
conjugacy classes  is
negligible relative to the standard deviation $\sigma_{\LH}$ of
$\Wos(\tau)$, and thus for our purposes we need only investigate the
statistics of $\Wos(\tau)$. 



\section{The mean and variance of $\Wos$}\label{mean and variance} 

\subsection{The averaging procedure} 
We define an averaging procedure by  taking 
a non-negative weight function $\weight\geq 0$, which is smooth and 
compactly supported in $(0,\infty)$,   
with $\intinf \weight(x)dx =1$.   
We then get  an averaging operator: 
$$
\aveT{F}:= \frac 1T \intinf F(\tau)\weight(\frac{\tau}\width)\d \tau \;.
$$
Let $\Prob_{\weight,T}$ be the associated probability measure:
$$\Prob_{\weight,T}(f\in \mathcal A) =
\frac{1}{T} \intinf \I_{\mathcal A}( f(t))
\weight\left(\frac{t}{T}\right)\;\d t \;.
$$

Note that the requirement $\weight \in C_c^\infty(0,\infty)$ implies
that the Fourier transform of $\weight$ decays rapidly:
$$ 
\^\weight(x) \ll |x|^{-A}, \mbox{ as } |x|\to \infty
$$
for all $A>1$. 
In the concluding section~\ref{sec:conc} we will relax the
 restrictions on $\weight$ to allow other averages, e.g.  
 $\weight=\mathbf 1_{[1,2]}$ so that we take $t$
uniformly distributed in $[T,2T]$, or  
 $\weight(t) = 2t \mathbf 1_{[1,\sqrt{2}]}$ 
when we take the eigenvalue $\lambda = 1/4 +t^2$ uniformly distributed
in $[E,2E]$, $E = 1/4+T^2$.

\subsection{The expected value of $\Wos$} 
We will first show that the mean value  $\aveT{\Wos }$ tends to zero
as $T\to \infty$ provided $\LH =O(\log T)$:  
Averaging \eqref{sc \Wos}  we find 
$$
\aveT{\Wos} = 
\frac 1{\pi \LH}\sum_{2<n<e^{\crad \LH}} \amp(n) 
\^f(\frac {\log \norm(n) }{2\pi \LH}) 
2\Re \^\weight(\frac{\width}{2\pi}\log \norm(n))  \;.
$$
Note that since $\log\norm(n) \sim 2\log n$ and $\supp \^f \subseteq
[-1,1]$,   the sum is over $n\leq e^{\crad \LH}$.  
Using  $\^\weight(x) \ll x^{-A}$ as $x\to \infty$, 
we have 
$$
\aveT{\Wos} \ll \frac 1\LH \sum_{n<e^{\crad \LH}} \amp(n)
\frac 1{(T\log n)^A}
$$
Since $\sum_{n \leq x} \amp(n)^2 \ll x$ by Lemma~\ref{Peter's
formula}, we have by the Cauchy-Schwartz inequality that 
$$
\aveT{\Wos} \ll \frac {e^{\crad \LH}}{T^A \LH^{A+1}}
$$
which goes to zero since we assume $\LH = O(\log T)$. 

Note that this argument also works when we allow straight averages
(such as $\weight = {\mathbf 1}_{[1,2]}$) as long as $\LH<\frac 1\pi \log
T$.


\subsection{The variance of $\Wos$} 

\begin{prop}
If $\limsup \crad \LH/\log T<1$ then as $T\to \infty$: 
\begin{equation}\label{formula for sigma L}
\aveT{ (\Wos)^2} \sim  \sigma_\LH^2 =: 
\frac{2\kappa}{\pi \LH} 
\int_0^\infty \^f(u)^2 e^{\pi \LH u} \d u  
\end{equation}
where $\kappa$ is given by \eqref{Peter's constant}. 
\end{prop} 

Note that we have
$$
e^{(1-\epsilon)\crad \LH /2} \ll  \sigma_\LH  \ll 
\frac {e^{\crad \LH/2} }{\LH} 
$$
for all $\epsilon>0$. 

\begin{proof}
To compute $\aveT{(\Wos)^2}$, use \eqref{sc \Wos} to get 
\begin{multline*}
\aveT{(\Wos)^2} = 
\frac 1{(\pi \LH)^2} \sum_{m,n<e^{\crad \LH}} \amp(m)\amp(n) 
\^f(\frac {\log \norm(m) }{2\pi \LH}) 
\^f(\frac {\log \norm(n) }{2\pi \LH}) \\
 \times 
\sum_{\epsilon_1,\epsilon_2=\pm 1} 
\^\weight( \frac{\width}{2\pi} (\epsilon_1 \log \norm(m) 
+\epsilon_2 \log \norm(n)) ) 
\end{multline*}

We now deduce that as $\width\to \infty$, the only non-vanishing contribution is from the
``diagonal terms''  where $\epsilon_1=-\epsilon_2$ and 
$$
\norm(m) =\norm(n)
$$ 
that is  $m=n$.



If $m\neq n$ we may use Lemma~\ref{use Liouville} to get a lower
bound
\begin{equation}\label{liouville for 2}
|\log \norm(m)  \pm  \log \norm(n) | \gg \frac 1{\min(m,n)} \;. 
\end{equation}
To be included in the sum, we need 
$ \norm(m), \norm(n)\leq  e^{2\crad \LH}$, that is $m,n \leq e^{\crad
\LH}$, and so 
$$
\^\weight( \frac{\width}{2\pi} (\epsilon_1 \log \norm(m) 
+\epsilon_2 \log \norm(n)) ) <<(\frac {\min(m,n) }{T})^A \ll
(\frac{e^{\crad \LH}}T)^A \;.
$$
Moreover, from $\sum_{n<x} \amp(n)^2 \ll x$ we get by Cauchy-Schwartz
that  the off-diagonal contribution is dominated by 
$$
\frac 1{\LH^2} (\frac{e^{\crad \LH}}T)^A \sum_{m,n<e^{\crad
\LH}}\amp(m) \amp(n) \ll 
\frac{e^{ A\crad \LH}}{\LH^2 T^A} e^{2\crad \LH}
$$
for all $A>1$. 
This  goes to zero if  $\pi \LH\leq (1-\delta)(\log T)$ for some
$\delta>0$, which we assume.

The diagonal terms $m=n$ give 
$$
\frac 1{( \pi \LH)^2} 
\sum_{n>2} \amp(n)^2 \^f(\frac {\log \norm(n) }{2\pi \LH})^2 
$$
(where we used $\^\weight(0) = 1$). Since there are two such terms
(corresponding to $\epsilon_1=-\epsilon_2=+1$ or $-1$),  we have
the total diagonal contribution being 
$$
2\frac 1{(\pi \LH)^2} 
\sum_{n>2} \amp(n)^2 \^f(\frac {\log \norm(n) }{2\pi \LH})^2 \;. 
$$
This  can be evaluated asymptotically as
$\LH\to\infty$ using  Peter's formula (Lemma~\ref{Peter's formula}) to give
\begin{equation*}
\frac{2\kappa}{\pi \LH} 
\int_0^\infty \^f(u)^2 e^{\pi \LH u} \d u  
 =: \sigma_{\LH}^2 \;. 
\end{equation*}
Thus we find 
$$
\aveT{(\Wos)^2} \sim  \sigma_\LH^2 
$$
if $\limsup \pi \LH/\log T<1 $. 
\end{proof}

\section{Higher moments} \label{sec:\Wos is gaussian} 
We can now show that $\Wos(\tau)$ has  Gaussian moments:  
\begin{thm}\label{thm:higher moments}
For $K\geq 3$ the $K$-th moment of $\Wos/\sigma_\LH$ converges to that
of a normal Gaussian provided that $\LH\to\infty$ with $T$ but that 
$\LH=o(\log T)$: 
$$
\lim_{T\to \infty} \aveT{ (\frac{\Wos(\tau)}{\sigma_\LH})^K } =  
\begin{cases}
\frac{(2k)!}{k!2^k} 
\;, & K=2k \text{ even}\\ 
\\
0 \;,& K \text{ odd}
\end{cases}
$$
\end{thm}

\subsection{Reduction to the pre-diagonal}  
By \eqref{sc \Wos} the $K$-th moments of $\Wos$ is given by 
\begin{multline}\label{Higher moment}
\aveT{(\Wos)^K} = \frac{1}{(\pi \LH)^K} \sum_{n_1,\dots,n_K<e^{\crad \LH}} \prod_{j=1}^K \amp(n_j) 
 \^f(\frac {\log \norm(n_j) }{2\pi \LH}) \\
\times 
\sum_{\sign_j=\pm 1} 
\^\weight(\frac{\width}{2\pi}(\sum_j \sign_j\log \norm(n_j))) 
\end{multline}

We now show that as  $\width\to \infty$, the only (possibly)
non-vanishing contribution to 
\eqref{Higher moment} is for terms satisfying: 
\begin{equation*} 
\sum_{j=1}^K \sign_j\log \norm(n_j) =0
\end{equation*}
that is we have 
\begin{multline} \label{sum epsilonK}
\aveT{(\Wos)^K} = \frac{1}{(\pi \LH)^K} 
\sum_{\sign_j=\pm 1} 
\sum_{ \sum_j \sign_j\log \norm(n_j) =0} 
\prod_{j=1}^K \amp(n_j)  \^f(\frac {\log \norm(n_j) }{2\pi \LH}) \\
+ O(\frac{e^{\alpha_K \LH}}{T^{\gamma_K}})
\end{multline}
for some $\alpha_K,\gamma_K>0$. Since $\LH=o(\log T)$ the remainder term
vanishes as $T\to \infty$. 

To prove this, recall that by  Lemma~\ref{use Liouville},  
if $\sum_{j=1}^K \sign_j\log \norm(n_j) \neq 0 $ then for some
$\delta_K>0$ 
$$
|\sum_{j=1}^K \sign_j\log \norm(n_j)|
\gg_K  \left(\prod_{j=1}^K \norm(n_j)\right)^{-2^{K-1}+1/2} \gg
e^{-\crad \delta_K \LH } 
$$
since only terms with $\norm(n_j)<e^{2\crad \LH}$  appear in 
\eqref{Higher moment}. Thus for these terms we have 
$$
\^\weight(\frac{\width}{2\pi}(\sum_j \sign_j\log \norm(n_j))) 
\ll (\frac {e^{\crad \LH\cdot \delta_K}} T)^A
$$
Replacing $\amp(n)$ by $\log^2n\ll \LH^2$ in \eqref{Higher moment} gives
that the contribution of the terms with $\sum_{j=1}^K \sign_j\log
\norm(n_j) \neq 0 $ is dominated by 
$$
\frac 1{\LH^K} \sum_{n_1,\dots, n_K <e^{\crad \LH}} \LH^{2K} 
(\frac {e^{\crad \LH\cdot \delta_K}} T)^A 
\ll \LH^K \frac{ e^{\crad \LH(K+ A\delta_K)}}{T^A} 
$$
Since $\LH=o(\log T)$, this vanishes as $T\to \infty$ (in fact we need
only assume that $\LH <c_K\log T$ for this, if $c_K$ is sufficiently small). 
This proves \eqref{sum epsilonK}. 

\subsection{Off-diagonal terms} 
In \eqref{sum epsilonK} we consider the sum of non-diagonal terms,
that is terms for which there is at least one index $j$ such that  
$n_j\neq n_i$ for all $i\neq j$. 
To handle these, we use Lemma~\ref{lem:Indep} which forces the
relation 
$$
\prod_{j=1}^K \norm(n_j)^{\sign_j}=1
$$
to decompose into a union of such relations.  Thus there is a
decomposition 
$$
\{1,2\dots, K\}=\coprod S_j
$$
so that in each  subset $S_j$ we have 
\begin{equation}\label{Ident S_j}
\prod_{i\in S_j} \norm(n_i)^{\sign_i} = 1
\end{equation}
and the norms $\norm(n_i)=(n_i+\sqrt{n_i^2-4})/2$ lie in the same real
quadratic field $\Q(\sqrt{d_j})$ for all $i\in S_j$. 
In the diagonal case there are $K/2$ such sets, e.g. $S_1=\{1,K/2+1\}$,
$S_2=\{2,K/2+2\},\dots$  and the identities are of the form 
$\norm(n_j)\norm(n_{K/2+j})^{-1}=1$, $j=1,\dots,K/2$.

In the off-diagonal case we assume that there
is a subset $S_j$ contains at least $3$ elements. 
The number $r$ of subsets is then at most $(K-1)/2$, since 
$$
K=\sum_{j=1}^r \#S_j \geq 3+ 2(r-1)
$$

To count such tuples of $n_i$, we denote for each subset
$S_j$ by $d_j$ the common value of the square-free kernel of
$n_i^2-4$, $i\in S_j$ and then write 
$$
n_i^2-4 = d_j f_{i}^2,\qquad i\in S_j
$$

Let $\epsilon(d_j)$ be the fundamental unit of the field
$\Q(\sqrt{d_j})$ and write $\norm(n_i) = \epsilon(d_j)^{2k_i}$, $i\in
S_j$. 
Since $\log \norm(n_i)\ll \LH$ we have $k_i\ll \LH/\log \epsilon(d_j)$,  
$i\in S_j$ and the relation \eqref{Ident S_j} 
implies $\sum_{i\in S_j} \pm k_i = 0$. Thus for each subset $S_j$
there are at most $O( (\LH/\log \epsilon(d_j))^{\# S_j-1} )$ solutions 
of \eqref{Ident S_j} with $\log \norm(n_i)\ll \LH$. 

Recall that we are summing over $\log \norm(n)\leq 2\crad \LH$. 
Using $\amp(n)\ll (\log n)^2\ll \LH^2$  
we find that the off-diagonal
contribution is bounded by the sum over all partitions
$\{1,\dots,K\}=\coprod_{j=1}^r S_j$ of 
\begin{equation}\label{sum partitions}
\LH^2 \prod_{j=1}^r \sum_{\epsilon(d_j)\leq e^{\crad \LH}} 
(\frac {\LH}{\log \epsilon(d_j)} )^{(\#S_j-1)/2}  
\ll \LH^K (\#\{d \text{ fundamental }:\epsilon(d)\leq e^{\crad \LH} \})^r
\end{equation}
where $r\leq (K-1)/2$ is the total number of subsets $S_j$ in our partition.

\begin{lem}
The number of fundamental discriminants $d>0$ for which
$\epsilon(d)<X$ is $O(X^{1+\delta})$ for all $\delta>0$.
\end{lem}
\begin{proof} 
We need to bound the number of fundamental discriminants $d$ for which
the fundamental solution $\epsilon(d) = (x_d+\sqrt{d}y_d)/2 $ of $x^2-dy^2 = 4$ is
at most $X$. Since $\epsilon(d)\sim x_d$, this is equivalent to
bounding the number of fundamental $d$'s for which $x_d\ll X$.  In
turn, this number is majorized by the number $\nu(X)$ of all triples $(d,x,y)$
of positive integers, with $d\equiv 0,1 \mod 4$,  
for which $x^2-dy^2 = 4$ and $x < X$, which is   
the sum 
$$
\nu(X) =\sum_{x<X} \# \{d,y\geq 1, d\equiv 0,1 \mod 4: dy^2 = x^2-4 \} \;.
$$
Since for $x\neq 2$ the number of pairs $(d,y)$ with $dy^2 =x^2-4$ is
at most the number of divisors $\tau(x^2-4)$ of $x^2-4$, we find that 
$$
\nu(X) \leq \sum_{2<x<X} \tau(x^2-4) \ll \sum_{2<x<X} x^\delta\ll X^{1+\delta}
$$
for all $\delta>0$, by virtue of the bound $\tau(n)\ll n^\delta$ for
all $\delta>0$. 

Note: A more refined argument  \cite[Lemma~4.2]{Sarnak82} shows
that $\nu(X)$ is asymptotic to $\frac{35}{16}X$, 
so that one can replace the bound $O(X^{1+\delta})$ by $O(X)$.  
\end{proof}

Thus we find that \eqref{sum partitions} is bounded by 
$$
\LH^K e^{(1+\delta)\crad \LH r}  \ll \LH^K e^{(1+\delta)\crad \LH (K-1)/2 }
$$
for all $\delta>0$. 
Since $\sigma_\LH\gg  e^{(1-\epsilon)\crad \LH/2}$
for all $\epsilon>0$, 
this shows that the sum of the off-diagonal terms is 
$O(\sigma_\LH^{K-1+\epsilon})$, for all $\epsilon>0$. 
To prove Theorem~\ref{thm:higher moments}  
it thus suffices to evaluate the diagonal contributions. 

\subsection{The diagonal contribution} 
Assume now that there is the same number of $+$ signs as there are $-$
signs. That is $K=2k$ is even, and there are $\binom{2k}{k}$ such
choices of signs. For simplicity assume the first $k$ are $+$ and the
last $k$ are $-$.  
Thus we have to evaluate the sum 
\begin{equation}\label{equal sum}
\frac{1}{(\pi \LH)^{2k}} 
\sum_{ \prod_{j=1}^k\norm(n_j)= \prod_{j=k+1}^{2k}\norm(n_j)} 
\prod_{j=1}^{2k} \amp(n_j) \^f(\frac {\log \norm(n_j) }{2\pi \LH})
\end{equation}

 There are $k!$ ways to pair off variables from the
first $k$ and the last $k$, such as the pairing $n_j=n_{k+j}$, $1\leq
j\leq k$. Each such pairing contributes a term 
$$
\frac{1}{(\pi \LH)^{2k}} \left(\sum_{n>2} \amp(n)^2  
\^f(\frac {\log \norm(n) }{2\pi \LH})^2 \right)^{k}  \sim 
 (\frac{\sigma_\LH^2}2)^{k} \;. 
$$

There are overlaps between the different ways of pairing off
variables, which correspond to intersection of diagonals such as
$n_1=n_2=n_3=n_4$. The contribution of these was already estimated in
the study of the non-diagonal terms, as they correspond to 
relations \eqref{Ident S_j} where some subset has more than two elements.  

Thus the total contribution of diagonal terms to
$\aveT{(\Wos)^{2k}}$ is asymptotically 
$$
\binom{2k}{k}\cdot k! \cdot
(\frac{\sigma_\LH^2}2)^{k} = 
\frac{(2k)!}{k!2^k} \sigma_\LH^{2k} \;. 
$$
This proves Theorem~\ref{thm:higher moments}. \qed

\section{Conclusion}\label{sec:conc}  
Since the Gaussian distribution is determined  by its moments, 
Theorem~\ref{thm:higher moments} implies 
\begin{thm}\label{main thm with weights}  
Assume that  $\LH\to \infty$ as $T\to \infty$ but $\LH=o(\log T)$. 
Then 
$$
\lim_{T\to \infty} 
\Prob_{\weight,T} (\frac{\Wf(\tau)-\meanf(\tau)}{\sigma_{\LH}} <x )  = 
\int_{-\infty}^x e^{-u^2/2} \frac{\d u}{\sqrt{2\pi}} 
$$
\end{thm}

So far we have assume that the weight function $\weight$ defining the
averages is in $C_c^\infty(0,\infty)$.  
To deduce the results for the standard averages 
($\weight = {\mathbf 1}_{[1,2]}$) as in Theorem~\ref{main thm}, one
proceeds by approximating ${\mathbf 1}_{[1,2]}$ by ``admissible''
$\weight$'s in a standard fashion, see e.g. \cite{HR}. The details are
as follows: 
Fix $\epsilon>0$, and approximate the indicator function
$\I_{[1,2]}$ above and below by smooth functions $\chi_{\pm}\geq
0$ so that $\chi_{-}\leq \I_{[1,2]} \leq \chi_{+}$, where both
$\chi_{\pm}$ and their Fourier transforms are smooth  and of rapid
decay, and so that their total masses are within $\epsilon$ of
unity: $| \int\chi_{\pm}(x)\d x-1 |<\epsilon$. Now set
$\omega_{\pm}:=\chi_{\pm}/\int \chi_{\pm}$. Then $\omega_{\pm}$
are  ``admissible''  and for all $t$,
\begin{equation}\label{approx}
(1-\epsilon)\omega_{-}(t) \leq \I_{[1,2]}(t) \leq (1+\epsilon)
\omega_{+}(t)
\end{equation}
Now
$$
\meas\left\{t\in[T,2T]\ :\ \frac{\Wos(\tau)}{\sigma_{\LH} } \in
\mathcal{A}\right\} =\int_{-\infty}^\infty
\I_{\mathcal{A}}\left(\frac{\Wos(\tau)}{\sigma_{\LH}}\right)
\I_{[1,2]}\left(\frac tT\right) dt
$$
and since \eqref{approx} holds, we find
\begin{multline*}
(1-\epsilon)
\Prob_{\omega_{-},T}\left\{\frac{\Wos(\tau)}{\sigma_{\LH} } 
\in\mathcal{A}\right\}
\leq \frac{1}{T} \meas\left\{t\in[T,2T]\ :\
\frac{\Wos(\tau)}{\sigma_{\LH} }  \in
\mathcal{A}\right\}\\
\leq (1+\epsilon) \Prob_{\omega_{+},T}
\left\{ \frac{\Wos(\tau)}{\sigma_{\LH} }
\in\mathcal{A}\right\}
\end{multline*}
By Theorem~\ref{main thm with weights} 
we find that 
$$
(1-\epsilon)\frac{1}{\sqrt{2\pi}}\int_{\mathcal{A}} e^{-x^2/2}
\;\d x  \leq \liminf_{T\to\infty} \frac{1}{T}
\meas\left\{t\in[T,2T]\ :\ \frac{\Wos(\tau)}{\sigma_{\LH} } \in
\mathcal{A}\right\}
$$ 
with a similar statement for limsup; since $\epsilon>0$ is
arbitrary this shows that the limit exists and equals
$$
\lim_{T\to\infty} \frac 1T \meas\left\{t\in[T,2T]\ :\
\frac{\Wos(\tau)}{\sigma_{\LH} } \in
\mathcal{A}\right\}  = \frac{1}{\sqrt{2\pi}}\int_{\mathcal{A}}
e^{-x^2/2} \;\d x
$$ 
which proves Theorem~\ref{main thm}. 

The same consideration applies to other positive statistics, such as
the number variance.

\end{document}